# Cloud Service ranking using Checkpoint based Load balancing in real time scheduling of Cloud Computing

Mohammad Riyaz Belgaum[1], Safeeullah Soomro[1], Zainab Alansari, Muhammad Alam[2,3]

[1] College of Computer Science, AMA International University, Kingdom of Bahrain.
[2] Institute of Business and Management (IoBM), Korangi Creek, Karachi, Pakistan
[3] Universiti Kuala Lumpur, (IPSUniKL), Kuala Lumpur, Malaysia
{bmdriyaz, s.soomro, zeinab}@amaiu.edu.bh, malam@(iobm.edu.pk,seu.edu.sa)

**Abstract.** Cloud computing has been gaining popularity in the recent years. Several studies are being proceeded to build cloud applications with exquisite quality based on user's demands. In achieving the same, one of the applied criteria is checkpoint based load balancing in real time scheduling through which suitable cloud service is chosen from a group of cloud services candidates. Valuable information can be collected to rank the services within this checkpoint based load balancing. In order to attain ranking, different services are needed to be invoked in the cloud, which is time-consuming and wastage of services invocation. To avoid the same, this paper proposes an algorithm for predicting the ranks of different cloud services by using the values from previously offered services.

## 1 Introduction

The term Cloud originated from the network diagram that shows the internet as a schematic cloud [1]. Cloud Computing being a network-based computing, connects heterogeneous systems in different types of the network like private, public and hybrid infrastructure. The applications and services can be accessed over the network and internet. The cloud computing resources like storage, processing, memory network bandwidth and virtual machines are efficiently managed by the Cloud Service Provider (CSP) [1], [2], [3] and [4] by making use of the available computing tools. Various Cloud Deployment models: [1] and [5].

Private cloud is restricted to one particular organization or institution. Moreover, the same organization or the third party is responsible for organizing and managing the cloud [1] and [5]. In Public Cloud, the term public makes it accessible to all through the cloud service provider. The service providers follow "pay-as-you-use" as the service of the resources are rendered by them. Regularly the services are offered to the users [1] and [5] without the knowledge of the resource location. Community Clouds have a comparable group of organizations functioning with the same interests to access the resources. The services are restricted to particularly those organizations in that group. One of the organizations in such a group will be responsible for organizing and managing the services [1] and [5]. In Hybrid Cloud the organizations use both private and public cloud based on the requirements. The services of such clouds are accessed using standardized interfaces [1] and [5].

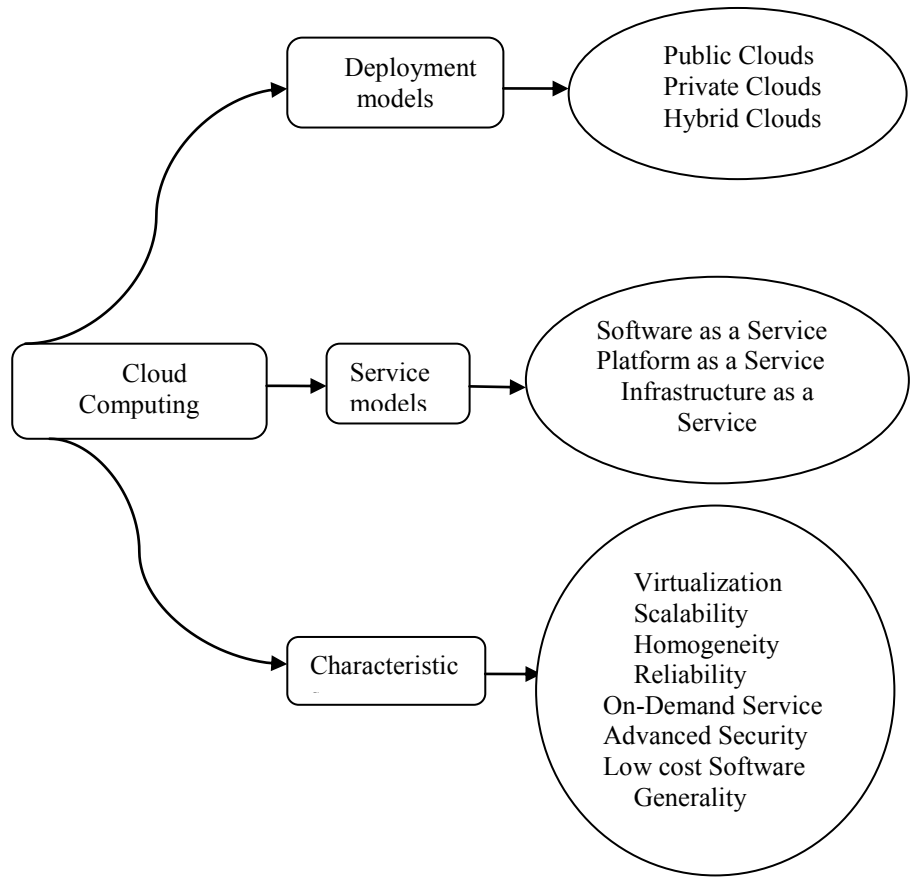

**Fig. 1.** Cloud Computing Overview.

There are various common characteristics of Cloud including massive scale, homogeneity, Virtualization, Resilient Computing, Low-cost software, Geographic distribution, service orientation and advanced security technologies [6]. Cloud services placed a number of challenges [7] to IT. Prominent of those are Service level agreement, Load balancing, security, the integrity of cloud services, costs, etc.

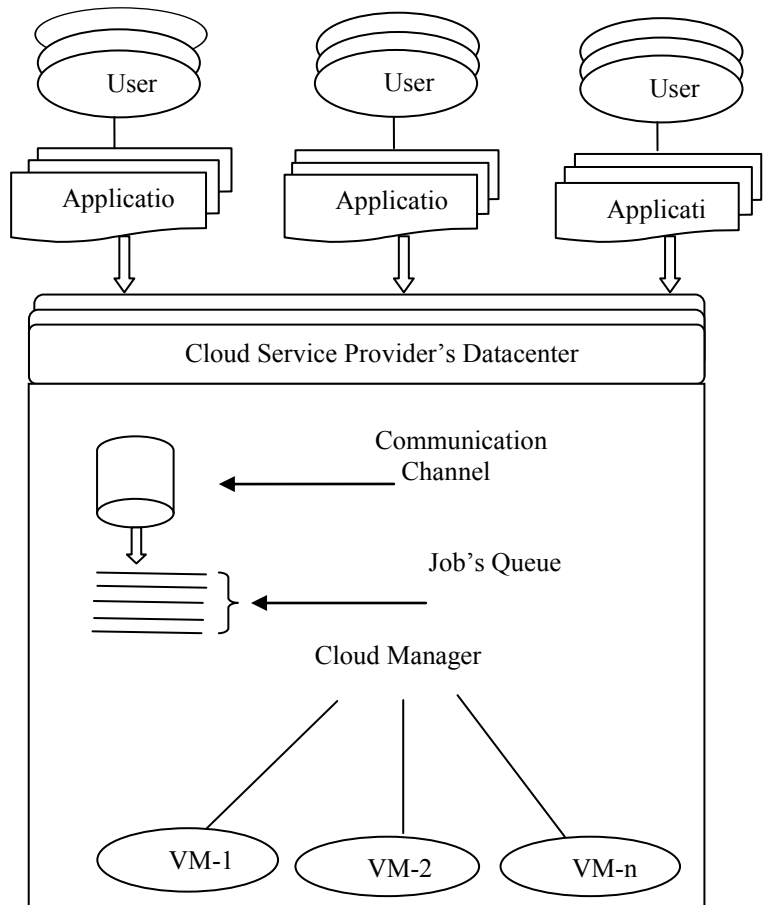

**Fig. 2.** Cloud Architecture for Load Balancing.

The checkpoint is a locally stable state of a process [8] which determines the consistency. Checkpoints are paramount and provide a point in time (PIT) copies of machines (VMs). Point in time (PIT) allows VMs to be rolled back to any point of the user's request. Checkpoints can be automatic and manual. Automatic checkpoints are taken regularly at every few seconds or as early as possible. These checkpoints are crash consistent and are useful during recovery. Manual checkpoints can be created by the users manually by allowing to control date and time. So it gives the user an option to do PIT recovery using manual checkpoints regularly. Executing the checkpoints helps to make the system be in a consistent state. And failure caused in any instance does not lead restarting of a process or to a long rollback. Checkpoints can be integrated with load balancing algorithms in order to provide high availability [9] to the requests of the clients in a no preemptive real-time scheduling environment.

The load balancer implements the virtualization technique [10] where the resources of the physical server at the data centers [11] are allocated to a wide range of virtual machines each of which performs the function of a physical server. Whenever the checkpointing occurs, the load of the physical server is analyzed and without preempting the resources from the available resources they are allocated to the virtual machine to accomplish the task. At any instance of time, the resources required or allocated must not be more than available.

## 2  Related Work

In [2], Pandey has followed multi agent system (MAS) in order to do the load balancing in Cloud Computing. MAS is a loosely coupled network of the software agents which work to solve issues which could not be solved using individual capacity. Authors used different parameters for load balancing which are reliability, configurability and ability to modify.

A checkpointing infrastructure for visualized service providers [1] was proposed by the authors. The authors used Another Union File system (AUFS) to bifurcate the read only from read and write parts in the virtual machine image. This idea almost works similar to the dirty bit (modified bit) architecture of the page replacement algorithm of Operating system. They have used Hadoop distributed file system for the check pointing architecture.

Cooperative Checkpointing discussed in [12] was compared with periodic checkpointing with an experimental analysis which shows the cooperative checkpointing helps the application or a process to proceed further under several group of failure distributions. Also the cooperative checkpointing is used to improve the reliability techniques like QoS, fault tolerance thereby making the system robust and increase the performance.

The authors proposed an algorithm [13] for reducing the execution time of tasks which are migrated from one VM to other VM when the deadline is not met by the task. Considering the metrics like throughput, profit, loss a simulation is carried out to show the non preemptive real time scheduling checkpointing decreases the execution time.

The Virtual Disk-Based Check-Restart [14] reduces the overhead of checkpointing by implementing a technique known as "Selective copy-on-write" taking the snapshots of the disks. Checkpoint-Restart proves to be more advantageous for the HPC applications as the execution resumes from the intermediate states and shows gain in the performance also reducing the bandwidth and storage.

For preventing the security threats in cloud computing, a design of internal traffic checkpoint model [3] was proposed. This model has three components which are used to identify and prevent threats to make the cloud resources secure.

The Load Balancing Techniques [15] along with the metrics are discussed to meet the challenges in the Cloud Computing. Different types of load balancing algorithms are categorized based on system load and system topology. Considering the first category the load can be centralized, distributed and also mixed. Where as in the second category as the topology can be static, dynamic and adaptive so are the load balancing algorithms.

The Identity-Based Authentication for Cloud Computing [16] focuses on the security and trust issue of the Cloud Computing. A new mathematical scheme based on encryption and signatures was proposed for the Hierarchical model for Cloud Computing. A comparative analysis was made between SSL Authentication Protocol and the proposed Identity-Based Hierarchical model for Cloud Computing (IBHMCC) Authentication protocol model to prove that the IBHMCC was more efficient and was certificate free.

The authors in [17] have proposed a new real time scheduling algorithm for cloud computing whose aim is to have a maximum utility of the resources using the time utility function. Two-time utility functions namely profit and penalty have been used in their work. The penalty was used to punish the tasks that have missed deadlines and the profit was used to reward the tasks which have met the deadlines.

The CPU Scheduling Algorithms in Cloud using CloudSim were analyzed by the authors in [18]. Using three of the CPU scheduling algorithms the results were simulated to show which will better suit the requirements of the user in the virtual environment of cloud. The virtual machine Manager decides to use which allocation policy and on to which virtual machine is it to be assigned. These experimental results help the cloud providers to charge the users based on their usage.

A preemptive cloud scheduling algorithm was used in [19] with a fixed priority assigned to every task in order to improve the QoS. Two variants of preemptive scheduling algorithms were mathematically proved to be fit for service oriented tasks. In these algorithms a dispatcher plays the role of preempting the low priority task when a high priority task arrives with less overhead and maintaining optimality to achieve QoS.

The authors in [20] made a survey of various load balancing algorithms in cloud computing presenting the challenges to be considered while the tasks were to be assigned to the Virtual machine in the cloud. The heuristic algorithms were proposed considering the load of the server as the load was monitored to disseminate the overload having a replica of the task at the machine where it was originated. The proposed approaches focused to improve the efficiency and satisfy the users.

In order to efficiently use the resources and to balance the load a genetic algorithm [21] was proposed by the authors. They classified the algorithms as static, dynamic and mixed scheduling algorithms. The methodology of considering the historical data and current state used calculates the load before it is deployed to a particular Virtual Machine. Using set theory, the relationship between physical machines and virtual machines was shown to know the load on each physical machine.

The authors in [22] considered load balancing as the main challenge along with achieving green computing with the various surveys. Because of the exponential increase of the cloud computing, the need for the data centers increased which in turn is resulting in excess of carbon emissions contaminating the environment. Various metrics to evaluate the load balancing algorithms along with Carbon emission and energy consumption metrics were used to show which algorithm is efficient.

The authors in [23] proposed an algorithm named "Load Balance Max-Min and max algorithm" for efficient load balancing in less completion time. A threshold is calculated to show the average completion time in all the machines. A case study is taken to make a comparison between the proposed algorithm and other algorithms to prove that the proposed algorithm completes the task in less time compared to others.

Load Balancing in conjunction with availability is discussed in [24] by the authors along with a Hospital Data Management system. In that case study the data of a patient needs to be accessed by different doctors, nurses globally from different systems when the information of the patient is available. A resource manager is responsible for the complete operations like monitoring, availability and performance.

## 3 Proposed Algorithm

The following section explains computation of the cloud service ranking using checkpoint based load balancing. As checkpoints can be either automatic or manual, here in our context we have considered automatic checkpoints as in manual based system, it can be fixed by the user. But in the automatic checkpoints, they take place regularly to have a proper load balancing in a real time environment. It is divided into two parts as Selection of services based on user's requirements with checkpoint based load balancing and the other ranking of corresponding services.

### 3.1 Service Selection based on user's requirements with checkpoint based load balancing:

The user will access the system based on the ranks obtained according to the services rendered before or have been accessed earlier. So initially, when all the jobs are submitted by the clients to the Cloud Service Provider (CSP), the CSP evaluates the load on each of the sub cloud and the number of times the jobs have been migrated from one sub cloud to another. And whenever the jobs are migrated from one sub cloud to other it is sent along with the previous saved checkpoint as the job has to resume from that point. In order to calculate correspondent services, rank at each sub cloud, a simple technique has been used. Based on the degree of correspondence between two services the ranking can be calculated considering checkpoint based load balancing according to user's requirements. For doing this, we will take the response time of a set of services with corresponding service nodes and then we will calculate the numbers of times these response times are inverted to convert one rank to other. The correspondence value of node is evaluated by

$$CV(\chi, y) = \frac{a - b}{n(n-1)/2}. \tag{1}$$

where n is total services, a is total number of consistent pairs and b is total number of variant pairs among two lists, $n(n-1)/2$ are the total number of pairs in the cloud with *n* services.

After evaluating the correspondence values among the node's service and the previously accessed similar services from the node, the corresponding node can be identified. For improving the accuracy of finding the corresponding node, we ignore the services with negative correspondence values and only include positive corresponding values.

Preferred nodes among the correspondent nodes are selected by subtracting the ranks of services,

(2)

$$P(\chi, y) = S_\chi - S_y.$$

Where $P(\chi, y)$ = prefer value among node x and y,

$S_\chi$ = rank of node x's service,

$S_y$ = rank of node y's service.

The greater prefer value indicates that the service is more reliable than the other service.

### 3.2 Ranking Services

The system then arranges the ranks of all the services in an order. Then the prefer values of each service is added with all the preferred values of the other services. The added value is priority value which means that the service at a particular node with greater priority value will be prioritized higher in the list.

$$PV = \sum_{y \in S} P(\chi, y). \qquad (3)$$

where $PV$ = priority value of service x.

The system then arranges the list having services with greater priority values higher in the list.

To improve the accuracy of rank prediction of services the system prefers the greater priority values of implicit services which the user has already accessed.

**Algorithm 1.** Proposed Algorithm

```
Input: A set of service S, x is a cloud service and π
stacks in the ranking.
 Output: ranked service list x
Step1: for each service from 1 to n
Step 2:         calculate correspondence value of each
     service based on user's requirements using eq. (1)
Step 3: end for
Step 4: for each service from 1 to n
Step 5: calculate prefer value of each service using eq.
     (2)
Step 6: end for
Step 7: R=S;
Step 8: for each service from 1 to n
Step 9: rank each service on the basis of checkpoints and
     the load balancing, present on the cloud,
               x = max rank in S,
               π(x) = S-R+1;
               R=R-x;
Step 10: end for
Step 11: for each service from 1 to n
```

```
Step 12:    calculate the priority value of each service
            using eq. (3)
Step 13:    rank   the   services   on   the   basis   of   their
            priority values,
                     R=μ(i)
                          a= max rank in priority value set,
            μ(i),
            π(x) = μ(i)-R+1,
            R=R-x,
Step 14: prioritize the implicit services with greater
                    rank.
Step 15:    update   the   service   set   S   with   the   ranked
                services and save it in ranked service list
                x
Step 16: end for
```

## 4   Conclusion and Future Work

In this paper, cloud service ranking has been done in checkpoint based load balancing in the real time scheduling. In this framework the user can access the services with higher rank by communicating with the users which have accessed the services in the past. By applying this technique, ranking of services can be improved and there will be less overhead in the cloud.

Simple ranking technique has been used in this framework. To improve the service ranking, other approaches can also be included that are based on rating the services. Other techniques can also be included to negate the checkpoint based load balancing in real time environment which reduce the prediction of services.

## References

1. Rejinpaul, N.R., Visuwasam M.L.: Check point based Intelligent Fault Tolerance for Cloud Service Providers. In: International Journal of Computers and Distributed Systems. Vol. 2, Issue. 1, pp. 59--64. (2012)
2. Pandey, R., Ranjan, R.MS.: Distributed Load Balancing in Cloud Computing. In: International Conference on Computer Science and Information Technology. pp. 32--36. (2013)
3. Eom, J.H., Park, M.W.: Design of Internal Traffic Checkpoint of Security Checkpoint Model in Cloud Computing. In: International Journal of Security and Its Applications. Vol. 7, No. 1, pp. 119--128. (2013)
4. Adari, V.R., Diwakar, Ch., Varma, P.S.: Cloud Computing with Service Oriented Architecture in Business Applications. In: International Journal of Computer Science and Technology. Vol. 3, Issue. 1, pp. 452--455. (2012)
5. Ahmed, M., Abu Sina, Md., Chowdhury, R., Ahmed, M., Rafee, M.H.: An Advanced Survey on Cloud Computing and State-of-the-Art Research Issues. In: International Journal of Computer Science Issues. Vol. 9, Issue. 1, No.1, pp. 201--207. (2012)
6. Okuhara, M., Shiozaki, T., Suzuki, T.: Security Architecture for Cloud Computing. In: FUJITSU Science and Technology Journal. Vol. 46, No. 4, pp. 397--402. (2010)

7. Buyya, R., Garg, S.K., Calheiros, R.N.: SLA-Oriented Resource Provisioning for Cloud Computing: Challenges, Architecture and Solutions. In: IEEE International Conference on Cloud and Service Computing. (2011)
8. Singh, D., Singh, J., Chhabra, A.: Evaluating Overheads of Integrated Multilevel Checkpointing Algorithms in Cloud Computing Environment. In: International Journal of Computer Network and Information Security. pp. 29--38. (2012)
9. Cao, N., Yu, S., Yang, Z., Lou, W., Hou, T.Y.: LT Codes-Based Secure and Reliable Cloud Storage Service. In: Proceedings IEEE INFOCOM. (2012)
10. Rashmi, K.S., Suma, V., Vaidehi, M.: Enhanced Load Balancing approach to avoid Deadlocks in Clouds. In: Special Issue of International Journal of Computer Applications on Advanced Computing and Communication technologies for HPC Applications. pp. 31--35. (2012)
11. Tsai, W.T., Sun, X., Balasooriya, J.: Service Oriented Cloud Computing Architecture. In: IEEE Seventh International Conference on Information Technology. pp. 684--689. (2010)
12. Oliner, A.J., Rudolph, L., Sahoo, R. K.: Cooperative Checkpointing: A Robust Approach to Large-Scale Systems Reliability. In: ICS 06: Proceedings of the 20th Annual International Conference on Supercomputing. pp. 14--23. (2006)
13. Santosh, R., Ravichandran, T.: Non Preemptive Realtime Scheduling using Checkpointing Algorithm for Cloud Computing. In: International Journal of Computer Applications. Vol. 80, No. 9. (2013)
14. Nicolae, B., Cappello, F.: BlobCR:Virtual Disk Based Checkpoint Restart for HPC Applications on Iaas Clouds. In: Journal of Parallel and Distributed Computing. 73, 5. (2013)
15. Sidhu, A.K., Kinger, S.: Analysis of Load Balancing Techniques in Cloud Computing. In: International Journal of Computers and Technology. Vol. 4, No. 2, pp. 737--741. (2013)
16. Hongwei, Li., Dai, Y., Tian, L., Yang, H.: Identity-Based Authentication for Cloud Computing. In: CloudCom. LNCS, 5931, pp. 157--166. (2009)
17. Liu, S., Quan, G., Ren, S.: Online Scheduling of Real-Time Services for Cloud Computing. In: 6th World Congress on Services. pp. 459--464. (2010)
18. Gahlawat, M., Sharma, P.: Analysis and Performance Assessment of CPU Scheduling Algorithms in Cloud using Cloud Sim. In: International Journal of Applied Information Systems. Vol. 5, No. 9, FCS, USA. pp. 5--8. (2013)
19. Dubey, S., Agrawal, S.: QoS Driven Task Scheduling in Cloud Computing. In: International Journal of Computer Applications Technology and Research. Vol. 2, Issue. 5, pp. 595--600. (2013)
20. Al Nuaimi, K., Mohamed, N., Al Nuami, M., Al-Jaroodi.: A Survey of Load Balancing in Cloud Computing: Challenges and Algorithms. In: Second Symposium on Network Cloud Computing and Applications. pp. 137--142. (2012)
21. Rawat, S., Bindal, U.: Effective Load Balancing in Cloud Computing using Genetic Algorithm. In: International Journal of Computer Science Engineering and Information Technology Research. Vol. 3, Issue. 4, pp. 91--98. (2013)
22. Kansal, N.J., Chan, I.: Cloud Load Balancing Techniques: A Step Towards Green Computing. In: International Journal of Computer Science Issues. Vol. 9, Issue. 1, pp. 238--246. (2012)
23. Hung, C.L., Wang, H., Hu, Y.C.: Efficient Load Balancing Algorithm for Cloud Computing Network. Supported by NSC, pp. 251--253.
24. Chaczko, Z., Mahadevan, V., Aslanzadeh, S., Mcdermid, C.: Availability and Load Balancing in Cloud Computing. In: International Conference on Computer and software modeling. Vol. 14, pp. 134--140. (2011)